# The structure of the *Arts & Humanities Citation Index*:

# A mapping on the basis of aggregated citations among 1,157 journals

Loet Leydesdorff,[a] Björn Hammarfelt,[b] and Alkim Almila Akdag Salah [c]

## Abstract

Using the *Arts & Humanities Citation Index* (A&HCI) 2008, we apply mapping techniques previously developed for mapping journal structures in the *Science* and *Social Science Citation Indices.* Citation relations among the 110,718 records were aggregated at the level of 1,157 journals specific to the A&HCI, and the journal structures are questioned on whether a cognitive structure can be reconstructed and visualized. Both cosine-normalization (bottom up) and factor analysis (top down) suggest a division into approximately twelve subsets. The relations among these subsets are explored using various visualization techniques. However, we were not able to retrieve this structure using the ISI Subject Categories, including the 25 categories which are specific to the A&HCI. We discuss options for validation such as against the categories of the Humanities Indicators of the American Academy of Arts and Sciences, the panel structure of the European Reference Index for the Humanities (ERIH), and compare our results with the curriculum organization of the Humanities Section of the College of Letters and Sciences of UCLA as an example of institutional organization.

[a] Amsterdam School of Communication Research (ASCoR), University of Amsterdam, Kloveniersburgwal 48, 1012 CX Amsterdam, The Netherlands; loet@leydesdorff.net ; http://www.leydesdorff.net .
[b] Department of Archival Science, Library and Information Science and Museology, Uppsala University, Thunbergsvägen 3H, 751 26 Uppsala, Sweden; bjorn.hammarfelt@abm.uu.se.
[c] The e-Humanities Group of the Netherlands Royal Academy of Arts and Sciences, Cruquiusweg 31, Amsterdam, the Netherlands; alelma@ucla.edu.

## 1. Introduction

Visualizing the information flow and the structure of sciences has become a tradition in the last decades of the 20th century. With the advancement of computer technology, and improved databases, new techniques are employed to generate the so-called *Science Maps* or *Atlas of* Sciences (Börner, 2010; Garfield, 1983; Small, 2003; Small & Garfield, 1985). Among these techniques, the methods of co-citation analysis (Marshakova, 1973; Small 1973; cf. Small & Sweeny, 1985), and aggregated co-citation relations among journals (Doreian, 1986; Doreian & Fararo, 1985; Leydesdorff, 1986, 1987; Tijssen *et al*., 1987) have been explored fully since the 1970s and 1980s, respectively. Whereas the outcomes of such studies have led to a novel understanding of the development of new research areas as well as the general structure of sciences and social sciences, all the maps created so far have had one important shortcoming: these maps either did not include the humanities or if included, the humanities were mostly discussed as an appendage to the social sciences.

A recent study by Klavans and Boyack (2009) provided an overview of twenty science maps: only four incorporated the humanities (based on the A&HCI dataset) and some did not even include the social sciences. In the consensus map of science, Klavans and Boyack (2009) attributed only a single field to the humanities among the 16 main areas of science. Thus, for example, "philosophy" and "art history" were considered as a single field. Mapping the A&HCI in more detail has perhaps suffered from the decision of the ISI (Thomson Reuters) not to produce a Journal Citation Reports for this index. They have been available for the other two indices (SCI and SSCI) since 1975 and 1978, respectively (Garfield, 1979).[1]

Garfield (1982b) explored the journal structure of the A&HCI, but he did not aggregate citations at the level of journals at the time (cf. Leydesdorff, 1994). Nederhof (2006), however, warned that the databases used for studying the social sciences and humanities have limitations (cf. Garfield, 1982a; Hellqvist, 2010, Hammarfelt, 2011; Linmans, 2010). Thus, the fine-structures of the humanities have been black-boxed and insufficiently unpacked; the available studies mainly focused on their positions relative to the social and natural sciences.

More recently, interest in the humanities has increased among policy makers and analysts (e.g., Aksnes & Sivertsen, 2009; Hicks, 2004; Hicks & Wang, 2009; Larivière *et al.*, 2006; Linmans, 2010). Using clickstream data, for example, Bollen *et al.* (2009) showed that the humanities and the social sciences are prominent when mapped from the perspective of the usage of scholarly journals. However, the humanities face a further complication: It cannot be distinguished clearly from the social sciences. Studies of social conflicts, for example, require a historical perspective, while "History" can primarily be considered as a humanities discipline. Furthermore, the emergence of transdisciplinary fields such as gender studies and cultural studies can blur the distinction between the social sciences and the humanities.

---

[1] The *Science Citation Index* has existed since 1961, but the first edition of *Journal Citation Reports* dates from 1975. The *Social Science Citation Index* was first published in 1966, and extended with *Journal Citation Reports* in 1978.

The use of the very term "humanities" to represent a set of disciplines was not spread before the 20th century—although the German universities used "Geisteswissenschaften" (mind sciences) during the 19th century. The disciplines within the humanities did not take their disciplinary and institutional forms in American universities until after WW I (Klein, 2005, p. 24). Nowadays, the humanities can be considered as one of three "cultures" within academia: the sciences, the social sciences, and the humanities (Kagan, 2009). Among these three, the social sciences are the youngest whereas the humanities house disciplines which were already part of the mediaeval universities.

Each of these three cultures has its own volume in the *Web of Science* database of ISI/Thomson Reuters. In this study, we focus on the A&HCI, which was launched in 1978 covering a little more than a thousand journals with the ambition to become an important tool for researchers in the humanities (Garfield, 1977). In a previous study, Leydesdorff & Salah (2010) analyzed the structure of the A&HCI from the perspective of a specific journal, namely *Leonardo*.

At the time, we downloaded the set of this database for the year 2008. In this follow-up study we raise the question of whether this data could be used for the comprehensive mapping of the arts and humanities as represented by the aggregated citations among journals and/or the ISI Subject Categories attributed to these journals (Rafols *et al.*, 2010)? Is a further integration of the three sets at the journal level feasible? Can a disciplinary structure in the A&HCI be reconstructed from this data? What would be a meaningful categorization of the humanities? Can the thus retrieved macro structure of the A&HCI be validated in comparison with other possible categorizations and visualizations of the humanities?

## 2. The intellectual and bibliographic organization of the humanities

The bibliometric approach to the humanities has been discredited by attempts to rank scholars, departments, and journals in these less codified areas with measures similar to those used in the natural and the social sciences. Journal rankings based on citation data and inclusion in the Web of Science have been criticized and bibliometrics has been discredited for being part of these administratively driven developments (Pontille & Torny, 2010, at p. 357). However, A&HCI—the main database in use for these rankings—has poor coverage of the humanities. In A&HCI only journal articles are processed, whereas books, book chapters, and other forms of communication actually account for most (> 70%; Larivière *et al*., 2006, at p. 1002) of the output of the humanities.

The practices of referencing in these scholarly domains are different from those in the (social) sciences, and furthermore they vary among the different specialties and disciplines that can be analyzed under the umbrella term of "humanities." Thus, the analysis of the journals covered by the A&HCI covers only a small part of the published literatures in these field. Most recently, Thomson Reuters announced the launch of a Book Citation Index later this year (Adams & Testa, 2011) and Google Books may provide options for future mapping attempts with a broader coverage (Kousha & Thelwall, 2009).

The revolt against the ranking exercise culminated in 2008 in a letter signed by sixty editors of journals in the category of "history and philosophy of science" who pleaded against the development of "initial" journal lists with classifications in terms of A, B, and C. These lists were initially proposed by the European Reference Index for the Humanities (*ERIH*), a project of the European Science Foundation (ESF). The editorial can be found, for example, at http://www.publish.csiro.au/?act=view_file&file_id=HRv19n2_ED.pdf (Howard, 2008).

On June 10, 2009, Scopus announced a further extension of their coverage of the humanities (to more than 3,500 journals) by using these "*ERIH*-lists." However, the foundation and the *ERIH*-project hastened to declare that the rankings were not meant as judgments of quality. Still, the need of these editors to defend the reputations of their journals against this intervention illustrates the sensitivity of the issue.

For example, *Leonardo: Art, Science, and Technology,* the leading journal in the field of experimentation in using new technologies in art (Leydesdorff & Salah, 2010; Salah & Salah, 2008) was classified on this list in the lowest category with a C. The unintended and unwarranted effects of such lists and rankings have been amply demonstrated (Howard, 2008; Laudel & Orrigi, 2006; Leydesdorff, 2008). Rafols *et al.* (2011), for example, were able to show that the top-lists of business schools can be strongly biased against more interdisciplinarily oriented units because of misplaced rankings in the journal lists.

The publication of the "*ERIH* lists" in a more final format has been postponed to May 2011. Recently, *ERIH* announced that it now operates with fifteen expert panels which are organized in fields as follows:

- Anthropology
- Archeology
- Art Architectural and Design History
- Classical studies
- Gender Studies
- History
- History and Philosophy of Science
- Linguistics
- Literature
- Musicology
- Pedagological and Educational Research
- Philosophy
- Psychology
- Religious Studies

**Table 1**: Fifteen fields distinguished in the *European Reference Index for the Humanities.* Source: http://www.esf.org/research-areas/humanities/erih-european-reference-index-for-the-humanities/erih-governance/erih-expert-panels.html

The same journals can be classified differently in categories across the panels.

We consider the worries of the sixty editors justified (Janik, 2011; Leydesdorff, 2008; Pontille & Torny, 2010). Intellectual organization is latently embedded in structures and these structures can differently be perceived and appreciated by agents and institutions given institutional interests (Leydesdorff, 2010). However, this study is not about ranking; we analyze structures in the organization of the humanities literature using statistical techniques. Rankings reduce these structures to hierarchies using a specific perspective.

The mere existence of the A&HCI for more than three decades covering and analyzing more than a thousand relevant journals can also be considered as a resource. As Garfield (1982a) noted, the structure of this database is surprisingly similar to those for the sciences and the social sciences, and there is no a priori reason for assuming that the latter two are less different from each other than either is from the A&HCI. In our opinion, the citation structure in the A&HCI can be considered as worthy of studying relatively independent of the use (or misuse) that can be made of such an analysis in research evaluations. We will return to these normative issues in the conclusions section.

At the other side of the Atlantic Ocean, the American Academy of Arts and Sciences recently developed a set of *Humanities Indicators* on the basis of a survey. The results of this project were presented in 2009, and are available at the Humanities Resources Center Online (at http://www.humanitiesindicators.org/). This study deliberately refrained from using bibliometric indicators, but collected data using a survey sent during the academic year 2007-2008 to 1,417 departments in humanities disciplines. The response rate was sufficiently high: 66%. Seventy-four indicators were organized into more than 200 tables and charts accompanied by essays. Eleven main topics were distinguished in the

"Statement of the Scope of the 'Humanities' for Purposes of the Humanities Indicators" (Table 2).[2]

| Discipline | Sub-fields |
|---|---|
| English Language and Literature | English, American, and Anglophone literature; general literature programs; creative writing; speech and rhetoric |
| Foreign Languages and Literatures | Modern languages and literature; linguistics; classics and ancient languages; comparative literature. |
| History | Includes history of science and medicine. |
| Philosophy | Includes history of philosophy. |
| Religion | Programs in the comparative, nonsectarian study of religion; studies of particular religions; history of religion; does not include programs in theology or ministry. |
| Ethnic, Gender, and Cultural Studies | Programs studying from an interdisciplinary perspective race, ethnic, gender, or cultural groups, such as Black studies, Hispanic studies, women"s studies, gender studies. |
| American Studies & Area Studies | Though some of these programs include strong social scientific components, their emphasis on history, language, and literature places them within the humanities. |
| Archeology | |
| Jurisprudence | Includes philosophy of law. |
| Selected Arts | Art history; the study of music, musicology, music theory and composition, and music history; the academic study of drama and cinema, but not programs primarily aimed at musical performance or music technologies. |
| Selected Interdisciplinary Studies | General humanities programs; programs in the study of a particular historical period (e.g., medieval and Renaissance studies, classical and ancient studies, holocaust studies, etc.). |

**Table 2:** 11 main topics of Humanities defined by the Humanities Resources Center Online at http://www.humanitiesindicators.org/statement.aspx.

The report added that "(t)he organizations and studies from which indicator data are drawn may include different disciplines within the humanities. For example, some count all theology and ministry courses as humanities instruction; others class history as one of the social sciences; still others assume all general education to be humanistic. […] Although political science, government, geography, anthropology, and sociology may, from certain perspectives, be considered *humanistic* social sciences, for the purposes of the Humanities Indicators, they are categorized as *non-humanities* disciplines. Interdisciplinary studies that link a predominantly social science perspective with humanities disciplines are also considered *non-humanities* studies."

[2] Statement of the Scope of the "Humanities" for Purposes of the Humanities Indicators. *Humanities Resource Center Online*. The American Academy of the Arts & Sciences (2010). Retrieved from http://www.humanitiesindicators.org/statement.aspx, retrieved on Feb. 1, 2011.

This distinction between "humanistic social sciences" and "humanities" can perhaps be operationalized in the A&HCI because in addition to the 1,157 source journals, approximately 1000 other journals are selectively introduced. These latter source items are mostly from the social sciences. The 1,157 core journals contained in the A&HCI are assigned to 66 ISI Subject Categories, whereas the larger set of 2,161 source journals are attributed to 167 of these 221 ISI Subject Categories. We shall focus on the 1,157 core journals exclusively attributed to the A&HCI in the journal analysis, but discuss now first the relation between these core journals and the "selectively introduced" other journals in terms of ISI Subject Categories.

Twenty five of these categories are different from the 221 ISI Subject Categories used for indexing the *Science* and *Social Science Citation Indices* 2008.[3] These 25 unique Subject Categories are listed in Table 2; they were attributed 1,022 times (in 2008) to the 1,157 source journals whereas a total of 41 (of the 221) other ISI Subject Categories were attributed 1,421 times to this set.

| *Subject categories* | *Frequency* |
|---|---|
| Archaeology | 57 |
| Architecture | 28 |
| Art | 56 |
| Asian Studies | 34 |
| Classics | 28 |
| Dance | 6 |
| Film, Radio, Television | 19 |
| Folklore | 15 |
| Humanities, Multidisciplinary | 82 |
| Language & Linguistics | 110 |
| Literary Reviews | 50 |
| Literary Theory & Criticism | 15 |

[3] In 2009, the number of ISI Subject Categories was extended with one to 222.

| | |
|---|---|
| Literature | 98 |
| Literature, African, Australian, Canadian | 6 |
| Literature, American | 14 |
| Literature, British Isles | 16 |
| Literature, German, Dutch, Scandinavian | 19 |
| Literature, Romance | 52 |
| Literature, Slavic | 9 |
| Medieval & Renaissance Studies | 25 |
| Music | 61 |
| Philosophy | 108 |
| Poetry | 13 |
| Religion | 80 |
| Theater | 21 |
| *Sum* | 1,022 |

**Table 3:** 25 Subject Categories specific to A&HCI.

In summary, Tables 1 to 3 provide three *a priori* categorizations of the humanities which are to a variable degree informed by bibliographic databases. The *Humanities Indicators* project emphasized that one did not wish to use these databases. The ERIH panels are informed by previous rounds of discussions with journal editors; the panels should provide legitimization to the more final lists which are to be published soon. The ISI Subject Categories are meant to facilitate information retrieval, but are neither literary warranted (like the catalogue of the Library of Congress) nor regularly updated (Bensman & Leydesdorff, 2009).

## 3. Methods and materials

Our data consists of a download in June 2009 of the records added to the A&HCI file between January 1 and December 31, 2008. This set contained 114,929 records attributed to the A&HCI on the basis of 2,161 source journals. As noted, approximately 1,000 of these journals were introduced selectively into the A&HCI in addition to the 1,157 sources that were fully covered by the A&HCI 2008.[4] These 1,157 core journals contain 110,718 of the 114,929 records (96.3%). We limit the analysis to these records because our focus is on the journal structure of the A&HCI itself.

We considered using Scopus data, but despite the larger number of journals currently contained in this database under the heading "arts and humanities" (1,935 journals), the number of documents covered by this database is still lower than in the A&HCI (Klavans & Boyack, 2009, at p. 464). Despite the realized extension to 1,935 journals, the retrieval for Scopus has remained smaller than that possible from the Web of Science (WoS).[5] At the time of this research (21 November 2010), the retrieval in 2010 for this subject area was 33,494 using Scopus versus 100,948 using the A&HCI at the WoS.[6]

A previous analysis of A&HCI data showed that the descriptive statistics of the ISI set remained consistent over the years. Leydesdorff & Salah (2010, at p. 791, Table 2) provided a breakdown of the document types and compared these numbers with the breakdown provided by Garfield (1982a, at p. 762) for the A&HCI in 1981, containing 101,362 records. The stability of the distributions was found significant ($\rho = 0.895$; $p < 0.01$). In Scopus, numbers have sharply risen over the last few years, in some cases by more than 20% per year. This confirms our earlier impression (Leydesdorff *et al.*, 2010) that the A&HCI is currently the more sophisticated available bibliographic database for the study of scholarly literature in the arts and humanities.

---

[4] Thomson Reuters lists 1,430 journals titles under the A&HCI at http://science.thomsonreuters.com/cgi-bin/jrnlst/jlresults.cgi?PC=H. However only 1,157 of these journal names matched records in the download for the year 2008.
[5] The Advanced Search option in Scopus provides the possibility to search with the subjarea() function. The help file indicates that one should use "subjarea(arts)" for retrieving the "arts and humanities."
[6] Klavans & Boyack (2009) concluded (at p. 463): "The Scopus database includes the majority of journals covered by TS (Thomson Scientific, L.), but adds a significant number of journals and proceedings from engineering, computer sciences, and health services." However, they added (at p. 464): "Scopus has very scant coverage of the humanities."

Leydesdorff & Rafols (2009) proposed a comprehensive map of the sciences and the social sciences using the ISI Subject Categories. These categories are attributed to the journals by indexers at Thomson Reuters for information retrieval purposes (Pudovkin & Garfield, 2002). Given this latter objective, most journals are assigned to more than a single category. However, these attributions may contain a lot of error from a scientometric perspective because they are based not on the analysis of citation patterns, but on library practices. Nevertheless, the attributions could be shown to provide a useful representation of structure because the error is averaged out at these high levels of aggregation in the case of the SCI and SSCI (Rafols *et al*., 2010; Rafols & Meyer, 2010).

The 25 ISI Subject Categories specifically used to classify journals in the humanities (Table 3 above) have a broad scope (e.g., "Humanities, multidisciplinary", "Philosophy" or "Religion") when compared to the Subject Categories used in the *SCI* and *SSCI* (e.g. "engineering, water" or "tropical medicine"). The latter are provided at the specialty level, while some of the ones in the humanities classify journals at the level of disciplines. The only field that seems to have warranted a more specialized categorization is "literature:" several categories are distinguished based on a specific language ("Slavic"), country or region ("British Isles" or "German, Dutch, Scandinavian") or genre ("Reviews" or "Poetry").

In summary, we use essentially two data matrices: the one of 1,157 aggregated journal-journal citations and the one of 66 Subject Categories based on aggregations in the journal matrix. From these two matrices, we will make selections such as when studying only the 25 x 25 matrix of Subject Categories that are unique for the A&HCI. The resulting matrices are factor-analyzed in order to explore their structure. We use the citing patterns throughout this study because "citing" is the running variable in each year, whereas "cited" is the sumtotal of citation to the archive of a journal (Leydesdorff, 1993a and b). The results of the factor analysis are used to color the partitions in the maps unless otherwise specified. Maps allow for the grasping of large structures ( > 1000 journals) easier than tables.

The maps are based on similarities between the citing patterns of the units of analysis (journals and/or Subject Categories). Although the factor analysis is first based on using the Pearson correlation coefficients between these variables, we use the cosine which has the advantage of not normalizing to the mean (Ahlgren *et al*., 2003). Particularly when distributions are skewed and with lots of zeros in the matrix this can improve the visualization. However, the cosine is defined from zero to one and threshold levels are needed, but there is not one-to-one correspondence between cosine values and values of the Pearson correlation (Egghe & Leydesdorff, 2009). We will select the visualization techniques pragmatically with the objective to generate maps of structures which can algorithmically be revealed in the data. The quality of these structures can be controlled, for example, in terms of so-called screeplots of eigenvalues using SPSS.

## 3. Results

### *3.1 Subject categories in the A&HCI*

Our initial ambition was to extend the set of 220+ ISI Subject Categories used for the mapping of the SCI and SSCI with the 25 specifically added for the A&HCI, and thus to integrate the A&HCI into the global base map of science provided at http://www.leydesdorff.net/overlaytoolkit. This base map can be overlayed with specific sets so that one can assess the disciplinary and/or interdisciplinary affiliations in any set downloaded from the Web of Science (Rafols *et al*., 2010).

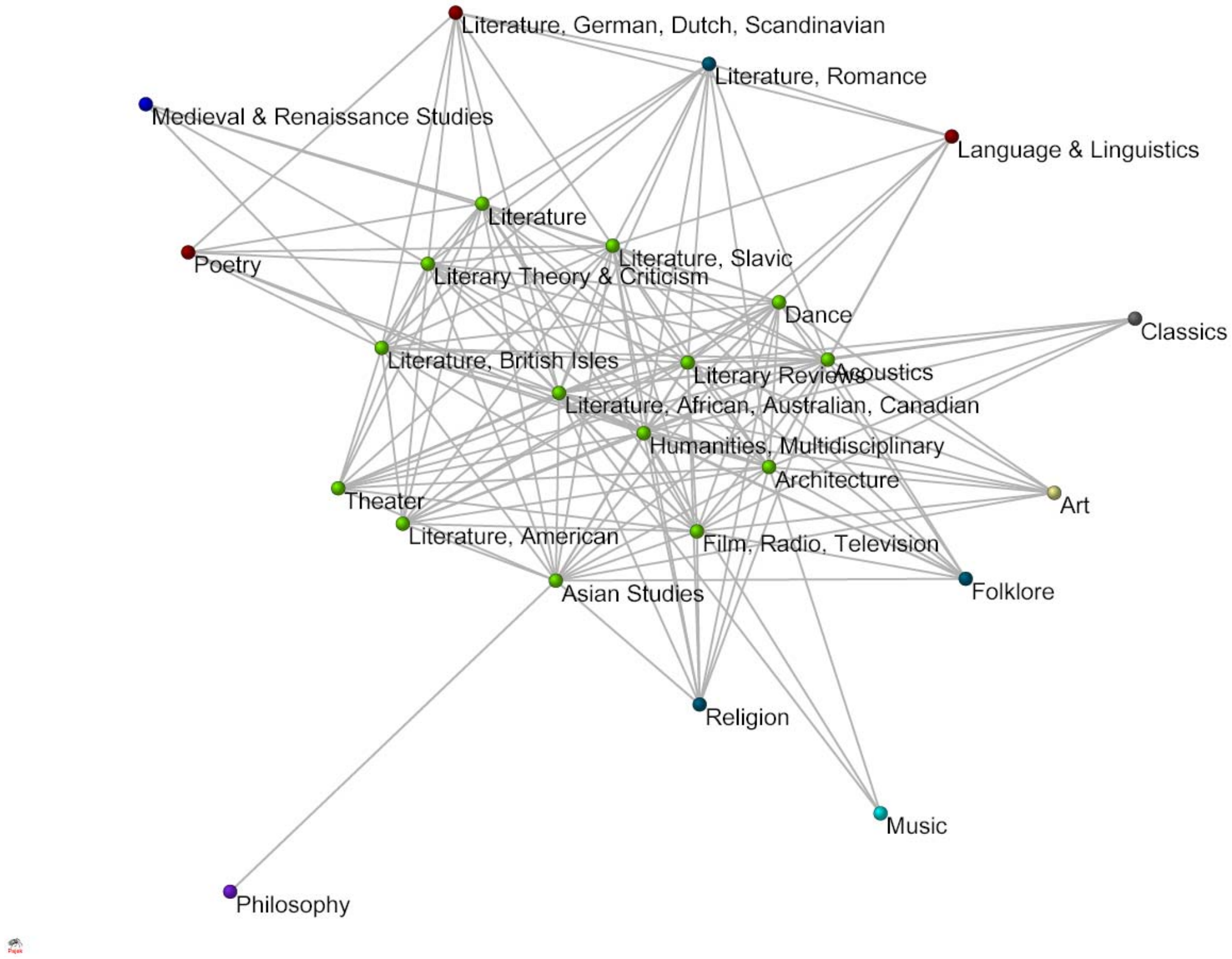

**Figure 1**: Cosine-normalize map of aggregated citation relations among twenty-five subject categories exclusive to the A&HCI; cosine > 0.5; Kamada & Kawai (1989).

This approach, however, did not work easily in the case of the A&HCI. Figure 1 first shows the map of the 25 ISI Subject Categories specific to the A&HCI. The set is partitioned (and accordingly coloured) using the *k*-core algorithm for the clustering. In this representation, the core cluster is not recognizable in terms of intellectual organization: “Architecture” and several branches of literature (e.g., “Literature, British Isles”; “Literature Slavic”) are grouped together, whereas “Literature, German, Dutch, Scandinavian”, for example, is mapped at a distance from this core set. Disciplines such as “Philosophy,” “Classics,” and “Language & Linguistics” are clearly separated from the core group. However, the single link between “Philosophy” and “Asian Studies” is not obvious. These links may be artifacts of low citation densities and incomplete indexing practices in the A&HCI (Larivière *et al*., 2006)

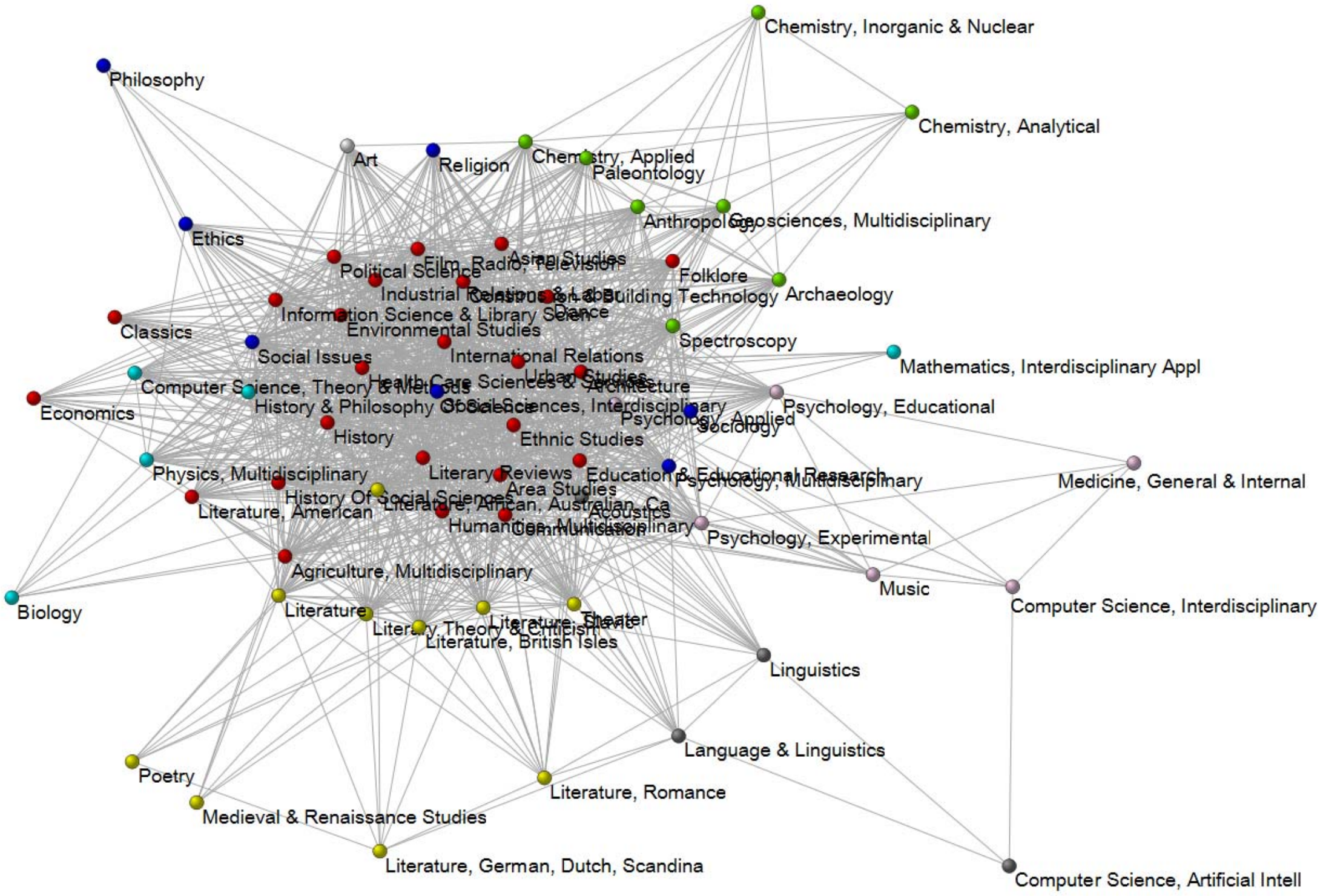

**Figure 2**: Sixty-six subject categories attributed 1,421 times to 1,157 journals in the A&HCI (Kamada & Kawai, 1989; cosine ≥ 0.5; seven factor solution used for the coloring).

In Figure 2, the 41 other subject categories (from the SCI and SSCI) assigned to these journals were also used for the mapping. Some links accord with expectations, such as the one between "Computer Science, Artificial Intelligence" and "Linguistics." Since the disciplinary organization is far from obvious, the coloring in the map was in this case based on factor analysis of the citation matrix. The factor analysis—based on the rotated factor matrix using the citing patterns of the aggregated journal sets as variables (cf. Leydesdorff & Rafols, 2009)—shows six other factors that represent groups other than the core one: archaeology (7.0%), philosophy (5.4%), literature (4.7%), psychology, music, and education (4.2%), "History and Philosophy of Science," and the other natural sciences (3.3%), and linguistics and artificial intelligence (2.8%). These seven factors explain 81.3% of the common variance. Art journals show interfactorial complexity on the first five factors.

Note that various subdisciplines of “Literature” such as “Literature, British Isles” and “Literature, Romance” are now organized in a separate grouping with “Poetry” and “Theater” whereas “Literature, American” and “Literature, African, Australian” have remained part of the core group. This core group is interwoven with social-science specialties which we classified above as “humanistically oriented” social sciences. Using this broader set, this integration can be made visible, but a clear structure cannot discerned.

Although relations in Figure 2 can thus be provided with meaning, the reasoning seems to remain a rationalization *ex post*. “Philosophy,” for example, fails to relate to various other disciplines—and is categorized differently from “History and Philosophy of Science”—while “Archaeology,” “Paleontology,” and “Anthropology” draw a number of chemical subdisciplines into their environment. The structure, however, shows how the humanities are deeply related to other disciplinary structures (Leydesdorff & Salah, 2010). The project of extending the Rafols *et al*.’s (2009) overlay with the humanities categories thus seems feasible using the full sets of journals and ISI Subject Categories for the three volumes of the Web of Science combined.

*3.2 The journal-journal matrix structure*

Let us now turn to the journal level. Figure 3 shows the cosine-normalized citation map colored using the main factor loadings on a twelve-factor solution. A zoomable map with journal labels (based on Gephi and Gefx; available at http://gephi.org/ and http://gexf.net/explorer, respectively) is available online at http://www.leydesdorff.net/A&HCI11/figure3.htm.

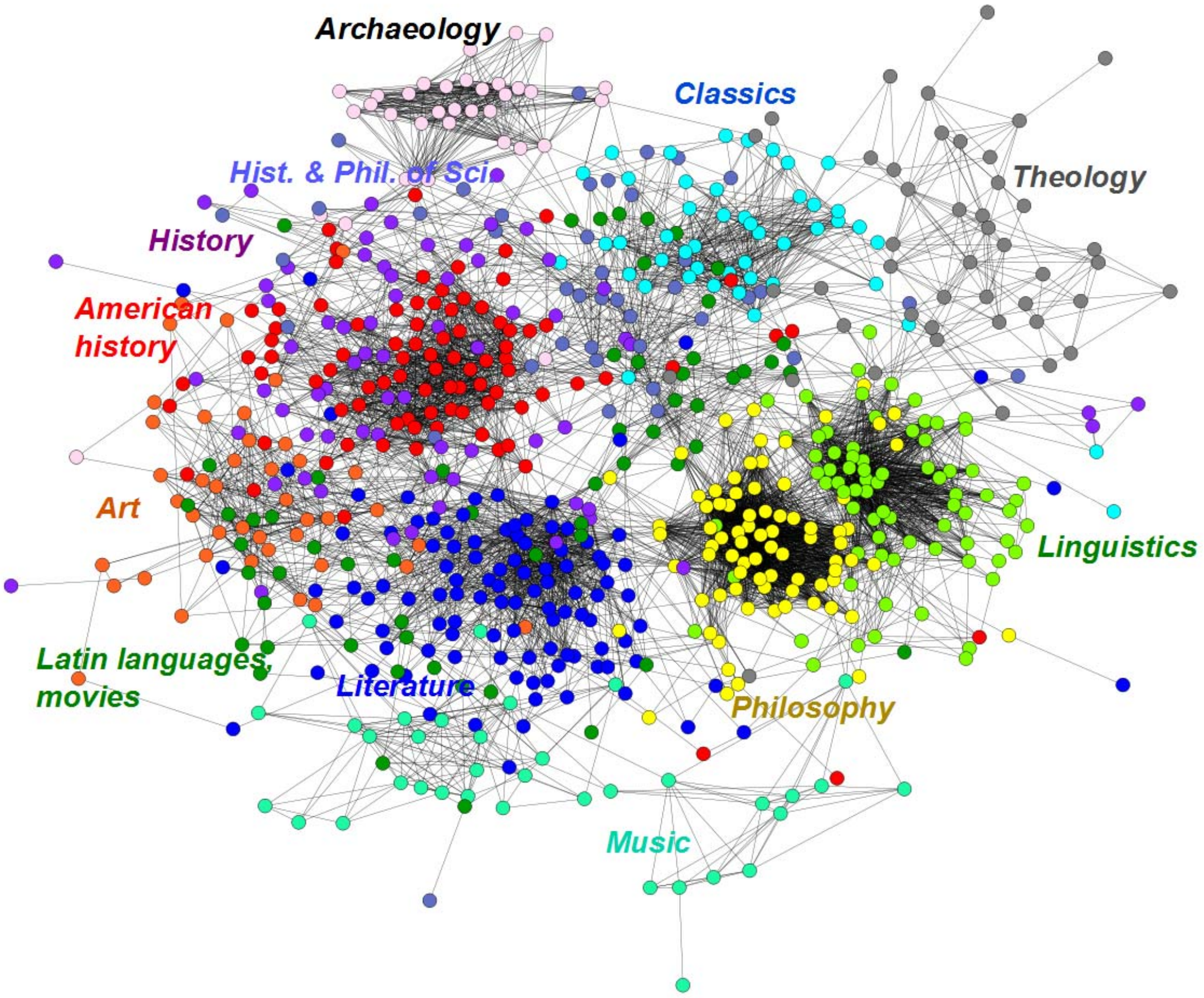

**Figure 3:** 724 journals related in their citing patterns with cosine ≥ 0.3 and with factor loading ≤ -0.1 or ≥ 0.1, colored in accordance with the 12 factor solution (Kamada & Kawai, 1989).

The choice for 12 factors (explaining only 18.4% of the variance) is a bit arbitrary given the gradual slope of the screeplot (based on the values of the eigenvalues in decreasing order), but the fit between the twelve categories designated on the basis of this factor solution and the figure was most convincing. With fewer than 12 factors, theology is no longer visible as a separate group. More than 12 factors leads to further fine-graining in the periphery of the graph. Allowing for 13 factors, for example, leads to a division within factor 11 into separate groups for Latin languages and movies. This division is meaningful, but the grouping of this factor is scattered in the visualization.

We labeled these 12 factors as follows (in Table 4).

| *Factor* | *Designation* | *% of Variance* | *Cumulative %* | *N* |
|---|---|---|---|---|
| 1 | Philosophy | 2.993 | 2.993 | 89 |
| 2 | Linguistics | 2.49 | 5.482 | 89 |
| 3 | American history | 2.059 | 7.541 | 100 |
| 4 | Literature | 1.938 | 9.479 | 119 |
| 5 | Archaeology | 1.503 | 10.981 | 47 |
| 6 | Classics | 1.466 | 12.447 | 54 |
| 7 | History | 1.295 | 13.742 | 81 |
| 8 | Art & art history | 1.122 | 14.864 | 52 |
| 9 | History & Phil. of Science | 1.005 | 15.869 | 45 |
| 10 | Music | 0.873 | 16.743 | 39 |
| 11 | Latin languages and movies | 0.872 | 17.614 | 66 |
| 12 | Religion | 0.822 | 18.436 | 52 |
| | *Other* | | | 274 |
| | *Total* | | | 1107 |

**Table 4:** Twelve factors distinguished by factor analysis (Varimax; SPSS v18.0).

Using the asymmetrical (2-mode) factor matrix directly as input to VOSViewer as another visualization program (available at http://www.VOSViewer.com/), an informative heat map of these twelve disciplines can be generated (Figure 4). One should note that VOSViewer uses a technique akin to multidimensional scaling for the layout of the map (Van Eck *et al*., 2010), whereas most other programs use spring-embedded algorithms (Kamada & Kawai, 1989). Therefore, the visualizations of these respective programs can be very different.

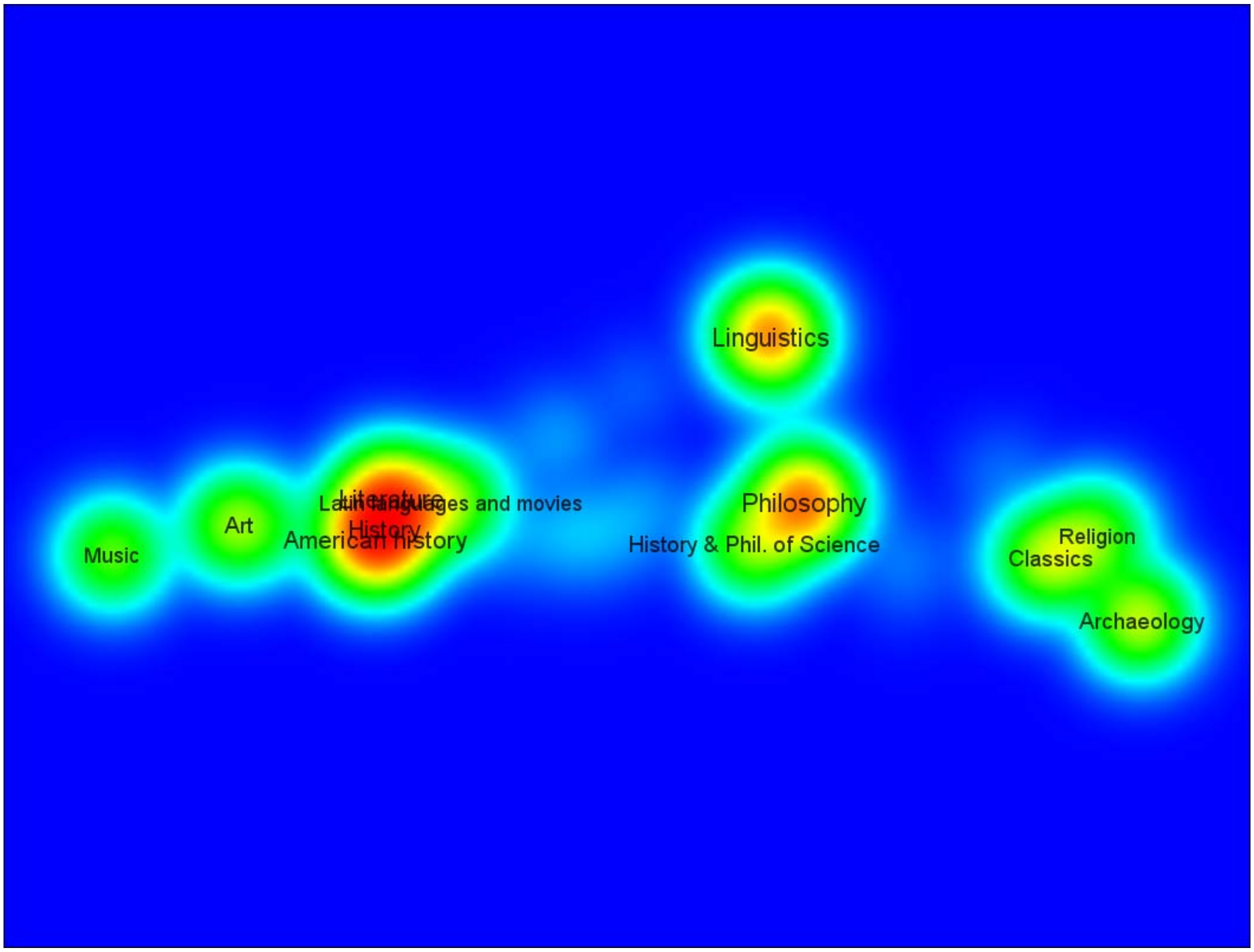

**Figure 4**: Visualization of the 12 main dimensions of 565 (of the 1157) journals included in the A&HCI 2008; factor loadings ≥ 0.2 or ≤ -0.2.

Although the labels of the categories "History," "American history," "Literature," and "Latin languages and movies" overlap in this visualization, the structure of the humanities is made very clear. "Linguistics" and "Philosophy," for example, can be considered as separate disciplines different from history and literature in terms of mutual citation relations. "Religion," "Classics," and "Archaeology" form a set of related disciplines (to the right in Figure 4), whereas "Art" (including art history) relates the domain of journals about "Music" (and music theory) with the major (and overlapping) fields of "History" (81 journals) and "Literature" (119 journals).

A file is brought online at http://www.leydesdorff.net/A&HCI11/figure4.htm which allows users to zoom into these twelve domains and retrieve the individual journals thus organized. Furthermore, the (cosine-normalized) citation environments of individual journals (listed in the A&HCI) were brought online in a gallery at http://vks2.virtualknowledgestudio.nl/A&HCI/browse.html and as Pajek input-files at http://www.leydesdorff.net/ah08/. The Pajek input-files were used for generating the gallery, but can also be used in other visualization and network analysis programs.

In summary, the delineation of the humanities into 12 subfields provides insight into the disciplinary organization of the A&HCI as a journal set. Major disciplines can be distinguished, such as "Philosophy," "Linguistics," "History," etc. One could perhaps argue for more or less refinement than twelve groupings, but the last factor of the twelve ("Religion"; 52 journals) made us decide to use a minimum of twelve fields. These twelve fields are less fine-grained than the 25 ISI Subject Categories which were specifically added to the A&HCI for the purpose of information retrieval.

## 4. Validation of the journal mapping of the A&HCI

We compared our results with two other categorizations of the intellectual organization of the humanities:

1. The US National Science Foundation maintains a database (at https://webcaspar.nsf.gov/) with information about earned doctorates (in the United States) with fine-grained disciplinary attributes. The table for the humanities was reanalyzed in the above mentioned project of the *Humanities Indicators.* This data provides us with a quantitative indicator of the distinctions among intellectual and disciplinary categories as made by the *Humanities Indicators*.

2. Following Balaban and Klein (2006), we mapped the network of departmental affiliations in the programs offered by the Humanities Section of the School of Letters and Sciences of the University of California at Los Angeles (UCLA). Our method diverged from Balaban and Klein (2006) in the choice of data; we used shared faculty among departments as connection points whereas their study was based on course requirements. UCLA can be considered as an example: its education programs are often interdisciplinary among departments.[7]

[7] For example, UCLA was ranked as number seven at http://www.topuniversities.com/university-rankings/world-university-rankings/2010/subject-rankings/arts-humanities. UCLA was ranked at the eleventh position at http://www.timeshighereducation.co.uk/world-university-rankings/2010-2011/arts-and-humanities.html.

The match between intellectual and institutional organization (at different levels) can be weak. Bourke & Butler (1998), for example, compared institutional information with field information as provided by the *Science Citation Index.* They concluded that "the interdisciplinary nature of modern scientific research, where researchers in departments publish in journals across a range of fields outside their nominal disciplinary affiliation, is an acknowledged 'norm' in the university research community."

One can expect the organization of faculties and departments and the intellectual organization to be even more uncoupled in the arts and humanities because of the already noted less institutionalized forms of specialization (Whitley, 2000). Furthermore, address information is often absent from the A&HCI so that bibliographic attribution of documents to the departments can be more difficult than in the case of the SSCI and SCI (Aksnes & Sivertsen, 2009; Nederhof, 2006; Larivière *et al*., 2006).

*4.1 PhD graduates*

Using the categorization provided in Table 2 (above), the *Humanities Indicators* project provided quantitative information about the number of doctoral graduates at http://www.humanitiesindicators.org/content/hrcoIIB.aspx. An Excel file with numbers is made available (at http://www.humanitiesindicators.org/binaries/II-11b.xls). This file allowed us to draw Figure 5 as a summary.

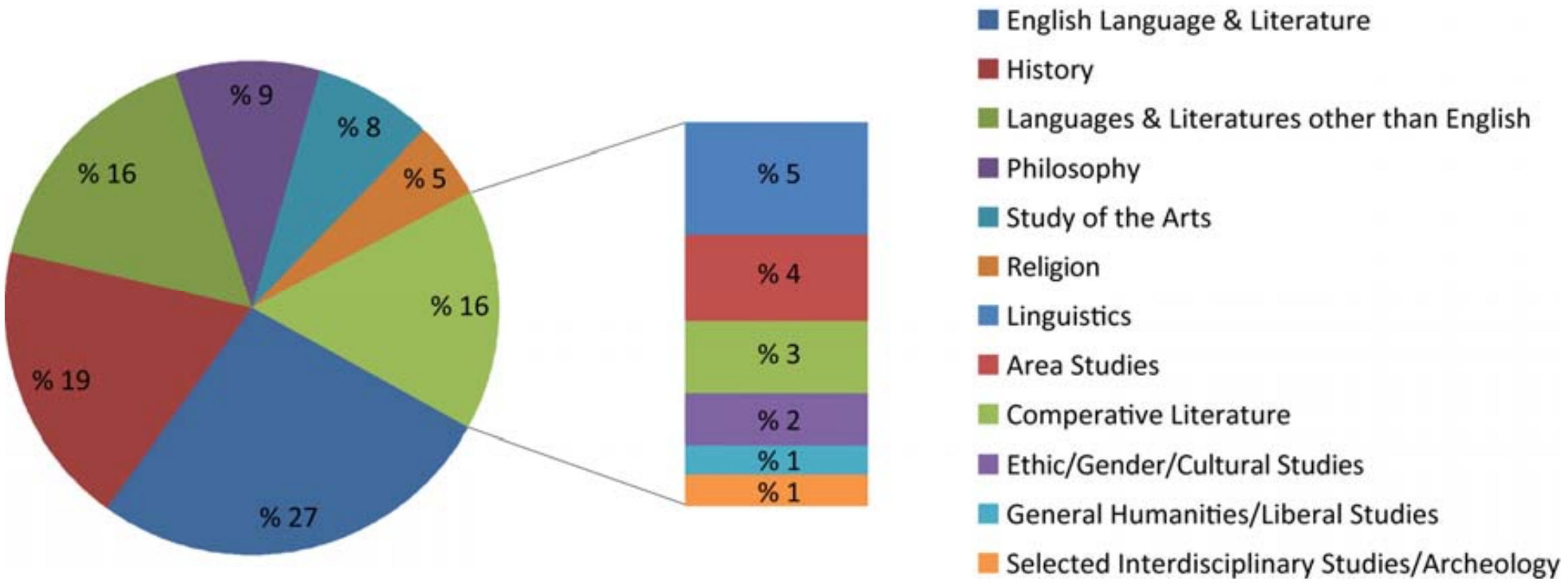

**Figure 5:** The distribution of Humanities PhD graduates in the USA in 2008. Source: U.S. Department of Education, Institute of Education Sciences, National Center for Education Statistics, Integrated Postsecondary Data System; accessed via the National Science Foundation's online integrated science and engineering resources data system, WebCASPAR (at https://webcaspar.nsf.gov/).

"Literature" can be considered as the backbone of the humanities (Klein 2005, p. 25). The prominence of studies in literature is evident both from the distribution of PhD graduates: English language & literature (27%), Languages and literatures other than English (16%), Comparative literature (16%), and in the number of literature journals that are indexed in the A&HCI (Table 2). The surprise in this table, however, is the position of "Linguistics" because it accounts for only 5% of the total number of PhD graduates in 2008, whereas it was the second largest dimension in the citation patterns of the A&HCI (89 journals). A grouping of such a size would be considered as a large group also in the SCI or the SSCI.

We note also the distinct position of linguistics in Figure 4 above. Linguistics thus is more prominently present in the journal literature than in the institutional organization. Georgas & Cullars (2005) noted that publication and citation practices in linguistics resemble those in the social sciences (e.g., sociology or economics) more than the other humanities. Actually, the major difference between our analysis and the listing in Table 2 is caused by the position of “linguistics.” Whereas we found “linguistics” to be a special group of 89 journals, it is classified under “foreign languages and other literatures” by the *Humanities Indicators*, that is, in the same group as “classics and ancient languages” (54 journals) and “modern literature.” The classification in this non-bibliometrically generated set of indicators thus remains in important respects puzzling.[8]

[8] The separate category of “Jurisprudence, including Philosophy of Law” is perhaps another case in point.

*4.2 Departmental structure in the humanities at UCLA*

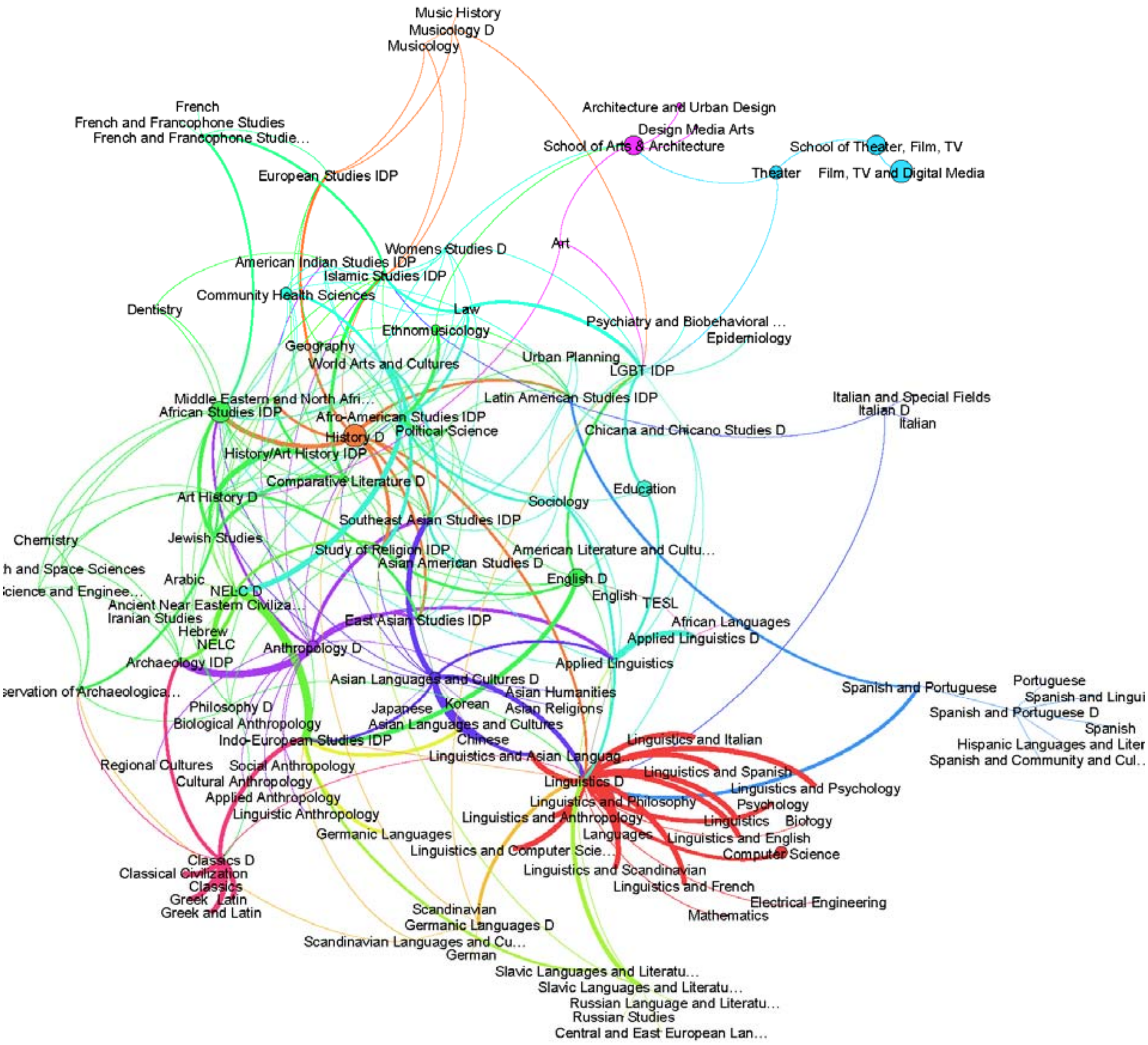

**Figure 6**: Departmental structure of UCLA Humanities section (according to shared faculty among teaching programs). Source: UCLA General Catalog 2010-2011, available at [http://www.registrar.ucla.edu/archive/catalog/2010-11/uclageneralcatalog10-11.pdf](http://www.registrar.ucla.edu/archive/catalog/2010-11/uclageneralcatalog10-11.pdf)

Figure 6 provides the network of (122) departments as linked by shared faculty in teaching programs in the Humanities Section of UCLA's School of Letters and Sciences. The network structure of approximately 300 such links draws History, Near-Eastern

Languages and Cultures, and English and Asian Studies in the center of the network. The section that is labeled as "Languages & literatures other than English" is scattered around the core. Italian, German, French, Spanish & Portugese, Scandinavian Languages & Literatures are positioned at the periphery of the network. Remarkably, French (at the top-left) and Portuguese (bottom-right) are placed at opposite ends although both are Latin languages. Perhaps, interdepartmental programs such as Lesbian, Gay, Bisexual and Transgender Studies, Women Studies, Afro-American Studies, and Indo-European Studies, etc., have changed the balance of the peripheries, and created a core that pushes the older and more traditional departments to the peripheries.

Note that the interdepartmental programs work with faculty from sometimes unexpected resources: the inclusion of Sociology, Education, Geography, and Political Sciences may not be surprising, but the presence of disciplines like Epidemiology or Dentistry is not so easy to explain. Chemistry, Material Sciences, and Earth and Space Sciences are linked to Archeology, and Psychology, Biology, Mathematics, Electrical Engineering, and Computer Science are part of the network of Linguistics. This orientation beyond the humanities accords with the selection that A&HCI makes from the wider literature, but these 4,000+ records (from appr. 1,000 journals) were not included in this study.

In summary, the organization at university level does not necessarily reflect the intellectual organization. Perhaps, more than in the natural and social sciences the humanities are evolving and fluid in their structures. Boundaries between intellectually different departments are systematically crossed in interdisciplinary programming since the subjects of scholarship are to be made relevant for audiences other than specialists sharing a common object or methodology (Klein, 2005). The institutional structures thus tell us little about the intellectual organization, whereas the textual (journal) citation structures do.

## Conclusions and discussion

The approach of the American Humanities Indicators project to neglect A&HCI (and Scopus as its younger companion), in our opinion, can be considered unreasonable. Survey data can also be biased and department heads may have reasons for defining the field as they do. Although an individual author may carefully select her references, the aggregated citation rates at journal level are beyond the control of individual agents and therefore more objectively determined according to the patterns of scholarly discourses.

The analysis of the aggregated citation relations in the A&HCI taught us that a latent structure in this data can be recognized in terms of an intellectual organization. One should always keep in mind that this is the citation structure in a specific domain, that is, in the literature insofar as it is published in the scholarly journals included in the A&HCI. However, these journals publish regularly, with established editorial boards, peer review, and other academic standards. Furthermore, the journals are monitored by the staff of Thomson Reuters, for example, in terms of their "quasi impact factors" (Stegman, 1999; Leydesdorff, 2008) although this citation data is not made available in a Journal Citation Report as in the case of the other two indices (Testa, 2003).

The major finding is that linguistics is much better and more coherently represented in this literature than it is in institutional survey data or in curriculum structures and numbers of PhD students graduating (cf. Georgas & Cullars, 2005). Some areas, such as linguistics and philosophy, can be considered as disciplinary structures in terms of the journals available. In other cases, it was more difficult to distinguish intellectual domains because of overlaps. For example, “history” and “literature” are intensively connected and so are to a lesser extent “religion,” “classics,” and “archaeology.”

A second finding is the relative absence of the humanistic social sciences from the core journals in the A&HCI database (Table 5). Gender, ethnic, and cultural studies are included in the Humanities Indicators and in the ERIH set, and they were pronouncedly visible from the curriculum analysis in Figure 6 above. As noted, articles from more than 1,000 journals are selectively introduced into the A&HCI by Thomson Reuters, but this additional set constituted only 3.7% of this database. Furthermore, we confined our analysis to the citation relations among the 1,157 journals which are fully covered by the A&HCI and this choice of journals may be conservative. As noted, the structure of the database has been relatively stable during the decades.

Factor analysis of the aggregated journal-journal citation matrix enabled us to sort the literatures contained in the database apart. Journal literatures about “art” and “music,” for example, could be indicated as two different domains. Let us again emphasize that a lot of creativity and scholarly production may be found at the margins of these different domains, for example, in books or journals not included in the A&HCI. It is not our intention to claim that this is *the* structure of the humanities. However, this journal structure is relevant in the humanities literature and it teaches us a lot about the organization of journals in the A&HCI. After this exercise one may hesitate to classify "linguistics" together with "other modern languages" and "classics" in a single category as it was done in the *Humanities Indicators* on the basis of survey data.

| *Factors (this study; Table 4)* | *ERIH (Table 1)* | *Humanities Indicators (Table 2)* |
|---|---|---|
| American history | Anthropology | American Studies & Area Studies |
| **Archaeology** | **Archeology** | **Archeology** |
| Art & Art history | Art Architectural and Design History | Selected Arts |
| Classics | Classical studies | |
| | Gender Studies | Ethnic, Gender and Cultural Studies |
| **History** | **History** | **History** |
| History & Philosophy of Science | History and Philosophy of Science | Selected Interdisciplinary Studies |
| | | Jurisprudence |
| Latin laguages and movies | | Foreign languages and literature |
| Linguistics | Linguistics | |
| Literature | Literature | English language and literature |
| Music | Musicology | |
| | Pedagological and Educational Research | |
| **Philosophy** | **Philosophy** | **Philosophy** |
| | Psychology | |
| **Religion** | **Religious Studies** | **Religion** |

**Table 5**: Organization of the humanities according to various classification schemes.

Table 5 organizes the classifications used in the *Humanities Indicators*, the *ERIH* project, and concluded by us on the basis of factor analysis of the journal-journal citation matrix, into a single scheme (cf. Klavans & Boyack, 2009). The three tables accord at four entries: archaeology, history, philosophy, and religion. In the case of "art & art history" the wording is different, but the same category may be meant. As noted, the *Humanities Indicators* subsumes some disciplines (e.g., linguistics) under larger categories. The *ERIH* project seems not to distinguish between "humanistic social science" and "humanities" as much as the other two projects did.

The other type of validation which we attempted in terms of departmental structures mainly made clear how institutional organization differs from intellectual organization. In a university context one may deliberately draw on varieties of expertise and bring scholars together from different backgrounds. The map of the UCLA departments based on shared faculty in the structures for organizing the education could not be matched to the intellectual organization retrieved at the level of journals. Universities develop historically along specific trajectories and of course with reference to intellectual environments, but this reflection is heavily mediated by administrative considerations and institutional incentives.

The fluent border between the cultures of the humanities and the social sciences is apparent from both the mappings in terms of the ISI Subject Categories (Figure 2 above) and in the departmental relations (Figure 6). The departmental network at UCLA can be considered as an example of the increasingly interdisciplinary nature of research and higher education (Klein, 2005). Therefore, it could be that the Subject Categories exclusively used in the A&HCI and the disciplinary labels used for the factor analysis in this study have failed to reflect an emerging structure of a "new" humanities in which transdisciplinary fields such as gender, ethnic, and environmental studies transgress borders between disciplines, and between the humanities and the social sciences. A further integration of the three databases in terms of aggregated journal-journal citations may enable us to reveal important transdisciplinary structures. Leydesdorff & Salah (2010), for example, found that art journals can be cited in journals that are classified as science and/or social science more than in the humanities.

The other major source that we had available were the ISI Subject Categories of which 25 were specifically developed for the A&HCI. ISI/Thomson Reuters keeps emphasizing that these categories are developed for information retrieval purposes and not as analytical categories; yet, they are often used in bibliometrics, for example, as indicators of interdisciplinarity (e.g., Morillo *et al*., 2001, 2003; Van Raan & Van Leeuwen, 2002). This interdisciplinarity, however, may find its origin in the overlap generated by dual or multiple attributions by the indexers of Thomson Reuters (Rafols & Leydesdorff, 2009). Using the other two databases, Rafols *et al*. (2010) showed that since these multiple attributions generate error that is not systematic, averaging at sufficient high levels may still lead to the revelation of useful structure among them. However, in the case of the A&HCI, we did not manage to shape a meaningful representation from the data at this level of aggregation. Perhaps the lower citation density makes random error stronger than the signal in this case.

In summary, journals more than the aggregated ISI Subject Categories were the relevant units of analysis for studying the latent structures in citation relations. The journals are very specific and grouped to an extent comparable to those in the *Science* and *Social Science Citation Indices* as Garfield (1982a) predicted. Maps depicting the environments of individual journals are also meaningful (Leydesdorff *et al*., 2010). The low citation rates do not prevent them from being very specific (Linmans, 2010).

**Acknowledgement**

We are grateful to Alexis Jacomy for adjusting the GEFXExplorer (at http://gexf.net/explorer/) to our needs. Björn Hammarfelt acknowledges the Swedish Foundation for International Cooperation in Research and Higher Education for financing his stay in Amsterdam.